\documentclass[preprint,authoryear,12pt]{elsarticle}

 \usepackage{graphics}

\usepackage{amssymb}

\journal{Comptes Rendus Geoscience}

\begin{document}

\begin{frontmatter}



\title{The effects of the overriding plate thermal state on the slab dip in an ocean-continent subduction system}


\author[geofisica1]{Manuel Roda}

\address[geofisica1]{Universit\`a degli Studi di Milano,
Dipartimento di Scienze della Terra ÒA. DesioÓ, Sezione di Geofisica, Via L. Cicognara 7, 20129 Milano, Italy.}

\author[geofisica1]{Anna Maria Marotta}

\author[geologia]{Maria Iole Spalla}

\address[geologia]{Universit\`a degli Studi di Milano,
Dipartimento di Scienze della Terra Ò A. DesioÓ Sezione di Geologia, and C.N.R.-I.D.P.A., Via Mangiagalli, 34, 20133, Milano, Italy.}

\begin{abstract}
To evaluate the effects of variations in the thermal state of the overriding plate on the slab dip in an ocean-continent subduction system, a 2-D finite element thermo-mechanical model was implemented. The lithosphere base was located at the depth of the 1600 K isotherm. Numerical simulations were performed while taking into account four different initial thicknesses for the oceanic lithosphere (60, 80, 95 and 110 km) and five different thicknesses of the overriding plate, as compared in terms of the continental-oceanic plate thickness ratio (100, 120, 140, 160 and 200\% of the oceanic lithosphere thickness). The results of numerical modeling indicate that a high variability of the subducting plate geometry occurs for an oceanic lithosphere thickness ranging from 60 to 80 km, while the variability decreases where the oceanic plates are thicker (95 and 110 km). Furthermore, the slab dip strongly depends on the thermal state of the overriding plate, and, in particular, the slab dip decreases with the increase in the upper plate thickness. The model predictions also confirm that a direct correlation between the slab dip and the age of the oceanic lithosphere does not exist, at least for subduction plates thinner that 110 km. These conclusions are supported by the good agreement between the model results and the natural data referring to worldwide ocean-continent subduction zones.
\end{abstract}

\begin{keyword}
Lithosphere thermal state \sep numerical models \sep slab dip \sep subduction zones

\end{keyword}

\end{frontmatter}


\section{Introduction}
\label{Introduction}
Several statistical analyses performed on subduction zones have indicated that a direct correlation between the slab dip and the age of the oceanic lithosphere does not exist \citep[e.g.,][]{Jarrard1986} and that other parameters may have a more significant impact on slab dip variations, including: (1) the absolute motion of the overriding plate \citep{Lallemand2005,Heuret2007}; (2) the subduction rate \citep{Cruciani2005,Lallemand2005}; (3) the nature of the overriding plate (oceanic or continental) \citep{Lallemand2005}; (4) the slab width \citep{Schellart2007, Guillaume2009}. Nevertheless, although the mutual interaction between these parameters controls the slab dip variations, the low cross-correlation coefficients obtained suggest that other parameters may have relevant roles \citep{Cruciani2005,Lallemand2005}.
In light of this, some analogue and numerical models were performed to investigate the influence of other causes of subduction geometry variation, such as: (5) the physical parameters of the slab  \citep{Royden2006}; (6) the back-arc stress variation \citep{Heuret2007,Clark2008}; (7) the slab interaction with the upper-lower mantle boundary in a long-term subduction \citep{Christensen1996,Olbertz1997,Guillaume2009}; (8) the trenchward motion \citep{Espurt2008}.
At present, no analysis has been performed to investigate the role of the thermal state of the overriding plate on the slab dip. To contribute to this subject, we used a 2-D finite element numerical model to develop a preliminary quantitative analysis of the impact of the thermal state of an ocean-continent convergent system. This kind of analysis could be very difficult if based solely on natural data. 
A comparison between the results of numerical modeling and the data obtained from natural systems allows the evaluation of the robustness and fidelity of the model. 
A convergence rate of 3 cm/a was chosen for the simulations. The primary motivation of this choice is that the present study is part of a wider analysis devoted to unravel the pre-collisional evolution of the Alpine belt. Within this geodynamic context 3 cm/a reveals the most appropriate subduction rate, as suggested by both geological evidences \citep{Agard2009} and parametric analysis \citep{Roda2010a}. On the other hand, while the effects induced by a variation of the subduction rate on the slab dip have been already widely investigated \citep[e.g.,][]{Cruciani2005,Lallemand2005,Roda2010a}, the goal of the present work is to enlighten the effects of a factor not yet taken into account, namely, the thermal state of the overriding plate.

\section{Model Setup}
\label{Setup}
We used the same approach as in \citet{Roda2010a}. The equation of continuity
\begin{equation} \label{eq:continuita}
\nabla á \cdot  \overrightarrow{v}=0
\end{equation}
momentum conservation
\begin{equation} \label{eq:momento}
\frac{\partial \tau_{ij}}{\partial x_{j}}=\frac{\partial p}{\partial x_{i}} - \rho \overrightarrow{g}
\end{equation}
and energy conservation
\begin{equation} \label{eq:energia}
\rho c \left( \frac{\partial T}{\partial t}+\overrightarrow{v} \cdot \nabla T \right)=-\nabla \cdot \left (-K \nabla T \right) + \rho H
\end{equation}
were solved by means of a finite element method within a 2-D rectangle domain (Fig. 1). In the equations  $\rho$ is the density, $ \overrightarrow{v}$ the velocity, $p$ the pressure, $ \overrightarrow{g}$ the gravity acceleration,
$\tau_{ij}$ the deviatoric stress tensor, $c$ the thermal capacity at constant pressure, $T$ the temperature, $K$ the thermal conductivity and $H$ the
heat production rate per unit mass. The integration of Equation (2) was performed using the penalty function formulation. The temporal integration of Equation (3) was performed using the upwind Petrov-Galerkin method with a fixed time-step of 0.1 Ma.

The model represents a strongly coupled ocean-continent convergent system \citep{Clift2004}, affected by slab dehydration and progressive mantle wedge serpentinisation; erosion and sedimentation processes are also accounted for, while shear heating is not taken into account  \citep{Marotta2007,Meda2010,Roda2010a}.
The initial geometry, the material parameters and the boundary conditions are the same as in \citet{Roda2010a} except for the convergent velocity of 3 cm/a, which here is fixed up to a 50-km depth in the trench zone, along a $45^o$ dip plane. The lithosphere base coincides with the 1600 K isotherm. 
We chose four different thicknesses of the oceanic lithosphere (60, 80, 95 and 110 km), and, to modify the thermal state of the overriding continental plate, we fixed for it five different thicknesses at a distance of 800 km from the trench, with a continent/ocean ratio ranging from 100\% to 200\% of the thickness of the oceanic lithosphere (Fig. 1 and Table 1).

\section{Model results}
In Figure 2, the results of the numerical simulations obtained for an 80-km-thick oceanic lithosphere are shown, after 25 Ma from the subduction start. The ratio of the continental/oceanic plate thickness ranges from 100\% (panel a1) to 200\% (panel a5) and a decrease of the mean slab dip with the increase of the upper lithosphere thickness is clearly evident (Fig. 2, from a1 to a5). 
Together with the slab dip decrease, a friction increase between the two plates occurs, increasing the ablation of the continental material belonging to the overriding plate, according to \citet{Faccenda2008} and \citet{Roda2010a}. 
In addition, in a low dip configuration, the efficacy of the crustal marker exhumation, driven by channel flow process, increases \citep{Roda2010a}, also producing the recycling of the upper oceanic crust and sediments (Fig. 2, panels a4 and a5). 
The large-scale stream-line analysis points out that the increase of the continental lithosphere thickness is accompanied by a reduction in the size of the right-side convective cell, until the development of a single-cell configuration (Fig. 2, b4 and b5). The large-scale left-side convective cell concurs with the buoyancy and the suction to the slab flattening. 

The influence of the lithospheric thickness variation (either oceanic and continental) on the subduction geometry was evaluated by plotting the slab-top profiles starting from a depth of 50 km, which is the lower limit of the fixed velocity vectors, and at 50 km from the trench, after 30 Ma from the subduction start (Fig. 3). Generally, the slab profiles change with changes in the lithospheric thickness of the overriding plates, and this variability increases with decreasing oceanic thickness.

For 60- and 80-km-thick oceanic lithospheres, the slab dip, evaluated to depth of 250-300 km, decreases with increasing thickness of the continental lithosphere (Fig. 3, panels a and b). Below a depth of 200 km, a gradual increase of the slab dip is predicted until reaching a near vertical setting followed, in some cases, by a dip inversion. For an oceanic lithosphere thicker than 80 km, lower dip variability is observed with the increase of the continental thickness (Fig. 3, panels c and d). This is due to the colder global thermal state of the system (because of the considerable plate thicknesses, either oceanic or continental), that drives the formation of a major single convective cell below the subducting plate and minimizes the intensity of the mantle circulation above the slab (Fig. 4, panels a and b), that is a primary controlling factor of the slab dip variability.

The analysis of the dip distribution from 50 to 400 km depth for oceanic lithospheres of 60 and 80 km and from 50 to 300 km depth for oceanic lithospheres of 95 and 110 km shows that a strong relationship exists between the slab dip and the thickness of the continental lithosphere (Fig. 5, panel a), with a rapid dip decrease with the continental plate thickness increase, in particular for the oceanic thicknesses of 60, 80 and 95 km. A low correlation between the slab dip and the continental thickness was observed, with a near constant dip between $27^o$ and $32^o$ for oceanic lithosphere thickness of 110 km.
The results obtained for the 80-km-thick oceanic lithosphere show the maximum slab dip variability.

The general dip decreasing with the increase of the continental lithosphere thickness also occurs when the imposed shallow angle is changed. The figure 6 (panels a to d) shows the slab dip variation for an entrance dip of $37^o$: in this case, the variability is less evident because of the higher efficiency of the overriding plate suction; the latter becomes dominant when a $30^o$ entrance dip is imposed, driving the underplating of the subducting plate and preventing the high dip variability with the thickness of the continental lithosphere (Fig. 6, panels e to h).

Although a direct relationship also exists between the slab dip and the continent-ocean thickness ratio, a weaker correlation is observed in comparison to the absolute thickness of the continental lithosphere (Fig. 5, panels a and b). This suggests that the thermal state of the overriding plate has more influence on the control of the slab dip than the variation of the continent-ocean thickness ratio.

In agreement with \citet{Jarrard1986}, \citet{Cruciani2005} and \citet{Lallemand2005}, the model predictions suggest that a direct correlation between the slab dip and the age of the oceanic lithosphere does not exist (Fig. 5, panel c).
\label{results}

\section{Comparison with the natural data}
To compare the model predictions and the natural data we considered the mean dips obtained for several worldwide ocean-continent subduction zones, extracted from the databases in \citet{Cruciani2005} and \citet{Lallemand2005}. The thickness of the overriding plate at a distance of 800 km from the trench for each natural data was based on the global thermal model of \citet{Artemieva2006}.
Due to the lacking of thermal predictions in the submerged areas, we excluded Japan and Aleutins subductions and some data on the Mexico, Philippines and Indonesia subductions, and we grouped the data into four classes of different oceanic lithosphere thickness: 60 km (data from 30 km to 70 km), 80 km (from 70 km to 85 km), 95 km (from 90 km to 100 km) and 110 km (from 100 km to 145 km). The thicknesses of the natural subduction plates derived as a function of the slab ages were obtained by \citet{Cruciani2005} and \citet{Lallemand2005} using the following relation \citep{Turcotte2002}:
\begin{equation} \label{eq:age}
y=2.32(kt)^{1/2}
\end{equation}
where $y$ is the thickness of the oceanic lithosphere, $k$ is the thermal diffusivity and $t$ is the slab age.
We used a total of 105 reference localities (Fig. 7, empty circles).

Even if the convergent velocities of the natural subduction systems were not discriminated, a good agreement between the natural data and the model predictions occurs, with a similar trend of the dip related to the thickness variation of the continental lithosphere: high slab dips seem to characterize the subduction systems with thinner overriding plates, and the dips decrease with increasing lithospheric thickness (Fig. 7). 
Furthermore, in natural cases, the maximum variability of the slab dip is associated with oceanic lithosphere thicknesses of 60 and 80 km, while lower variability characterizes the thicker oceanic plate thickness (e.g., 110 km), in agreement with the results obtained from numerical simulations (Fig. 5, panel a, and Fig. 7).
\label{natural}

\section{Conclusions}
The model predictions suggest that the slab dip is influenced by the thermal state of the overriding plate, and, in particular, for the cold overriding plate configuration, the efficacy of the right-side convective cell decreases and a shallower slab dip setting occurs. In contrast, in the hot overriding plate configuration, a highly efficient convective flow is developed in the right side of the model domain and a steeper subduction occurs. Furthermore, a thickness increase of the subducting plate decreases the variability of the slab geometry, with the development of a stable single-convective cell, disregarding the thermal state of the overriding plate.
 
The model results also indicate that a direct correlation between the slab dip and the age of the oceanic lithosphere does not exist, at least for subduction plates thinner that 110 km, as suggested by the statistical analysis of natural subductions. On the contrary, our comparative analysis strengths the correlation between the slab dip and the thermal state of the overriding plate.

Therefore, this study indicates that a factor not yet taken into account, namely, the thermal state of the overriding plate, exerts a relevant influence on the subduction geometry, although other simulations with different convergent rates are needed to enforce the cross-correlation between the different factors and to quantify the role played by each one.
\label{conclusions}

\section{Acknowledgements} 
We would like to thank the anonymous reviewers for their constructive criticism of the text. Prin 2008 "Tectonic trajectories of subducted lithosphere in the Alpine collisional orogen from structure, metamorphism and lithostratigraphy",  C.N.R.-I.D.P.A. and PUR 2008 "La ricerca geofisica: esplorazione, monitoraggio, elaborazione e modellazione" are gratefully acknowledged.
\label{Acknowledgements}

\begin{table}
\caption{List of the performed numerical simulations (O.L., oceanic lithosphere; C.L., continental lithosphere).}
\begin{tabular}{crrcr}
\hline
\hline
Model & O.L. & C.L.  & C.L./O.L.\\
 ID           &    thickness (km)             &  thickness (km)            &  ratio \\
\hline
01&60&60&100\%\\
02&60&70&120\%\\
03&60&85&140\%\\
04&60&95&160\%\\
05&60&120&200\%\\
\hline
06&80&80&100\%\\
07&80&95&120\%\\
08&80&110&140\%\\
09&80&130&160\%\\
10&80&160&200\%\\
\hline
11&95&95&100\%\\
12& 95&115&120\%\\
13& 95&130&140\%\\
14& 95&150&160\%\\
15& 95&190&200\%\\
\hline
16&110&110&100\%\\
17& 110&130&120\%\\
18& 110&150&140\%\\
19& 110&175&160\%\\
20& 110&220&200\%\\
\hline
\hline
\end{tabular}
\end{table}

\begin{figure}
\noindent\includegraphics[width=20pc]{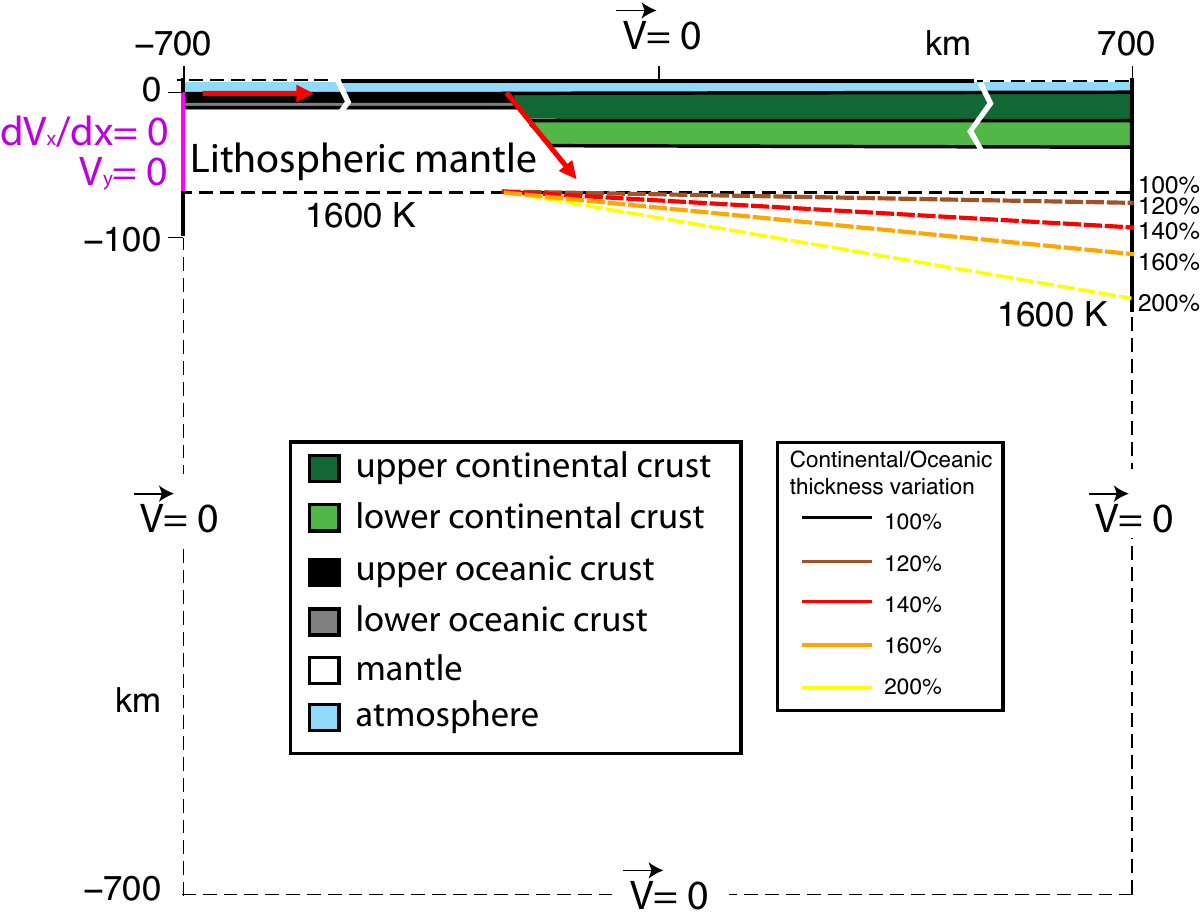}
\caption{Model setup and boundary conditions for all of the simulations. The legend illustrates the marker provenance. The different fixed depths of the 1600 K are also displayed. Distances are not in scale.}
\end{figure}

\begin{figure}
\noindent\includegraphics[width=20pc]{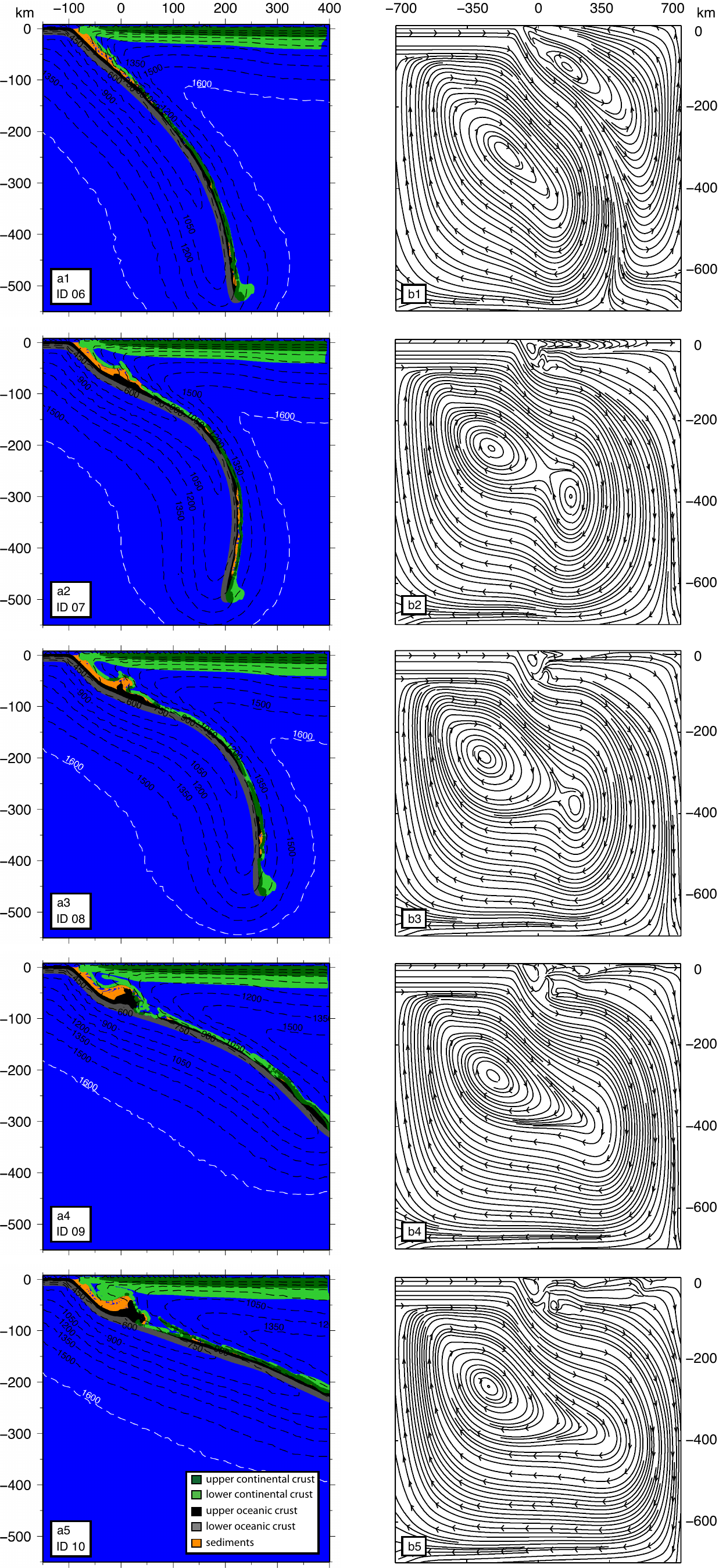}
\caption{Simulation results obtained for 80 km thick oceanic lithosphere and for 5 different thermal states of the overriding plate (from 100\% (panel a1) to 200\% (panel a5) of the thickness of the oceanic lithosphere), after 25 Ma since the beginning of the subduction. The white isotherm represents the base of the lithosphere (1600 K). In the right side (panels b) the stream-lines for each simulations are shown.}
\end{figure}

\begin{figure}
\noindent\includegraphics[width=39pc]{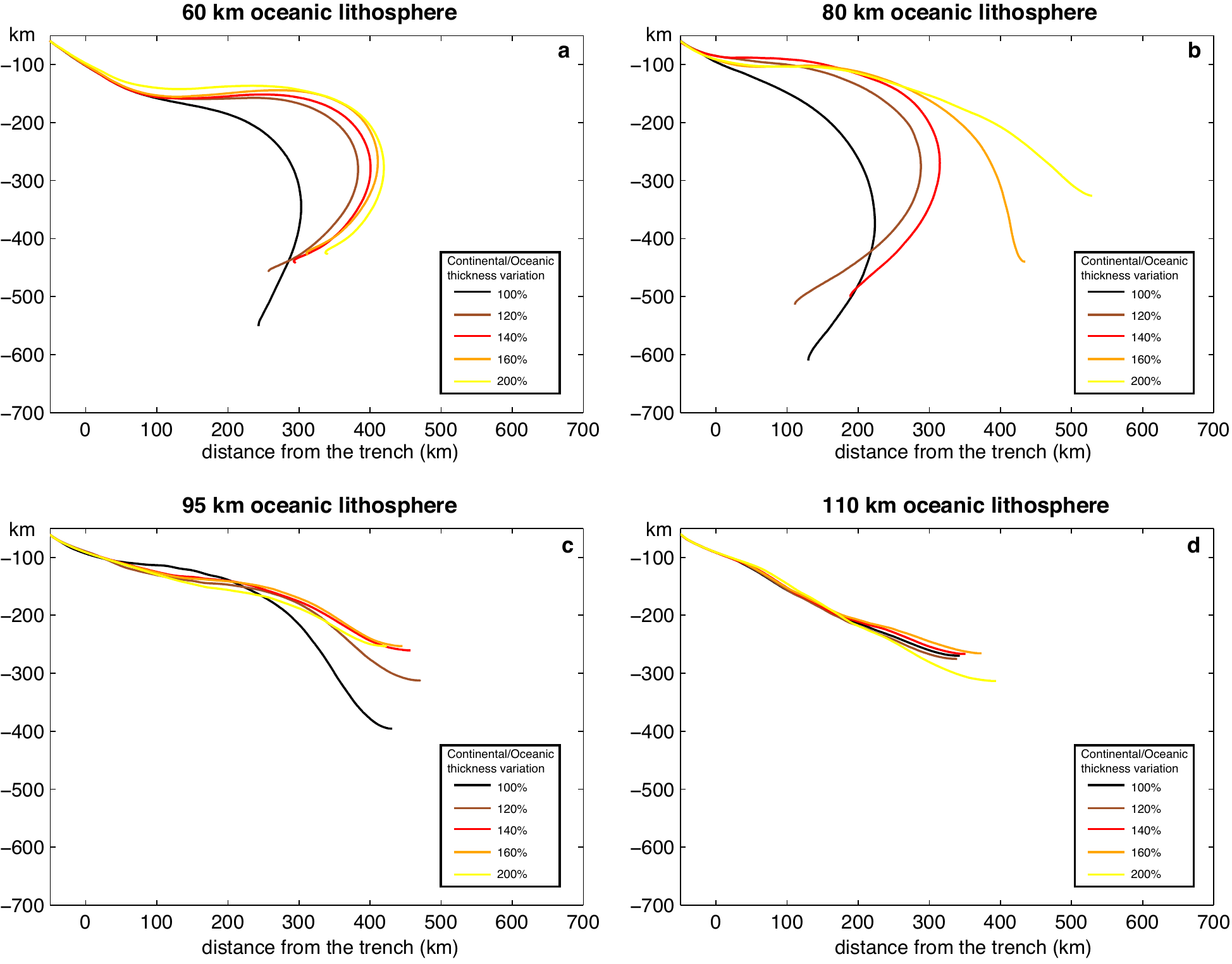}
\caption{Traces of the slab top obtained starting from 50 km depth for each performed simulation, after 30 Ma since the beginning of the subduction. Different colors discriminate the initial thicknesses of the overriding plate, as shown in the legend.}
\end{figure}

\begin{figure}
\noindent\includegraphics[width=20pc]{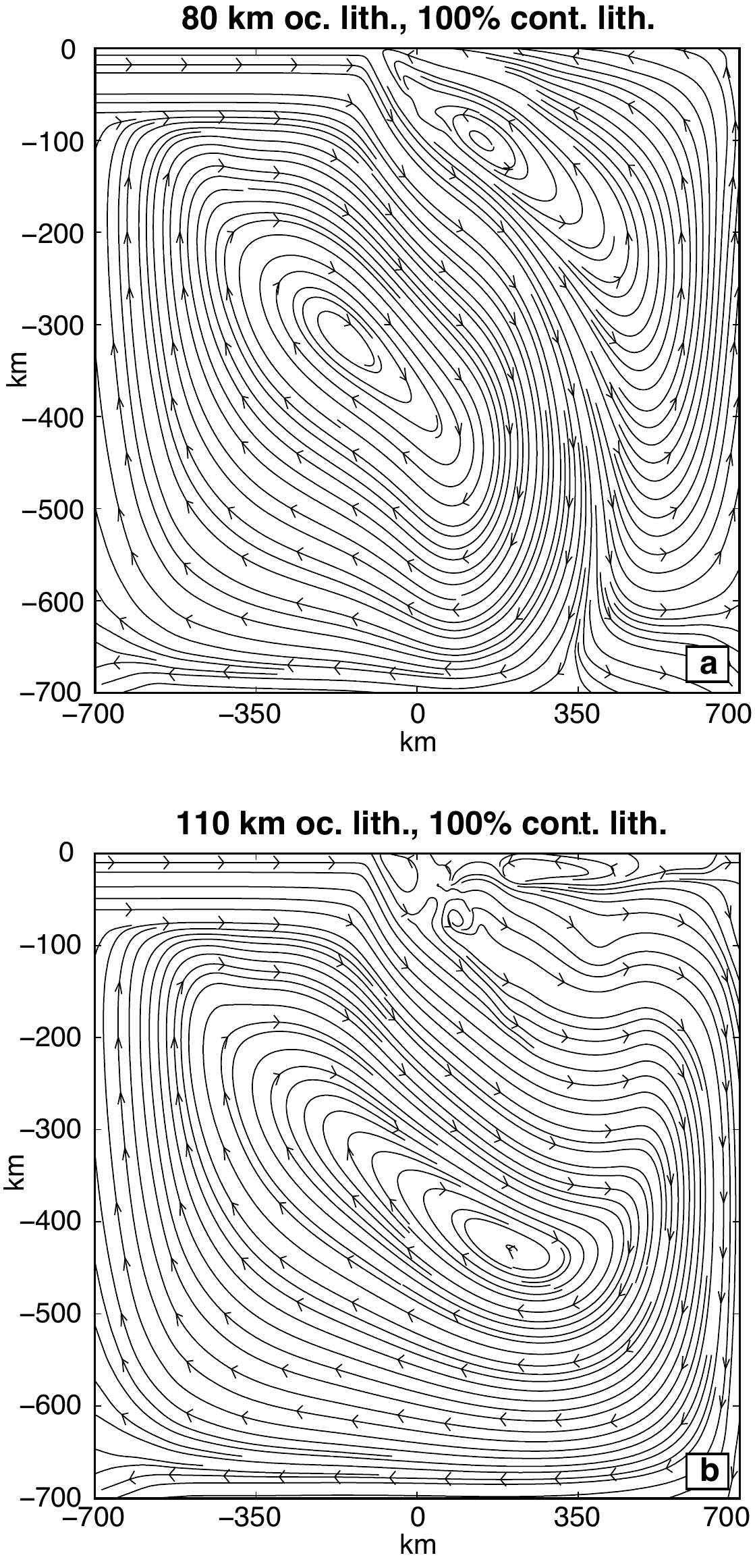}
\caption{Stream-lines obtained for 2 different oceanic thicknesses (80 km, a and 110 km, b) at the same continental/oceanic thickness ratio (100\%).}
\end{figure}

\begin{figure}
\noindent\includegraphics[width=16pc]{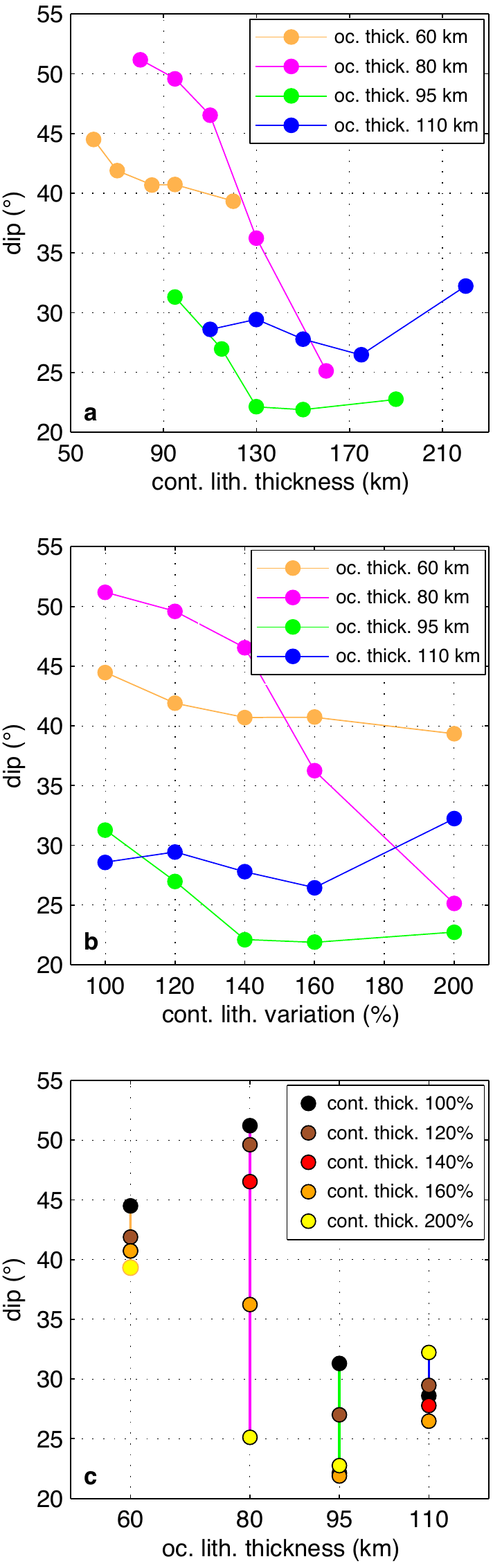}
\caption{Statistical analysis of the results of the performed simulations: panel a, mean slab dip vs thickness of the continental lithosphere; panel b, mean slab dip vs continental/oceanic thickness ratio; panel c, mean slab dip vs thickness of the oceanic lithosphere.}
\end{figure}

\begin{figure}
\noindent\includegraphics[width=20pc]{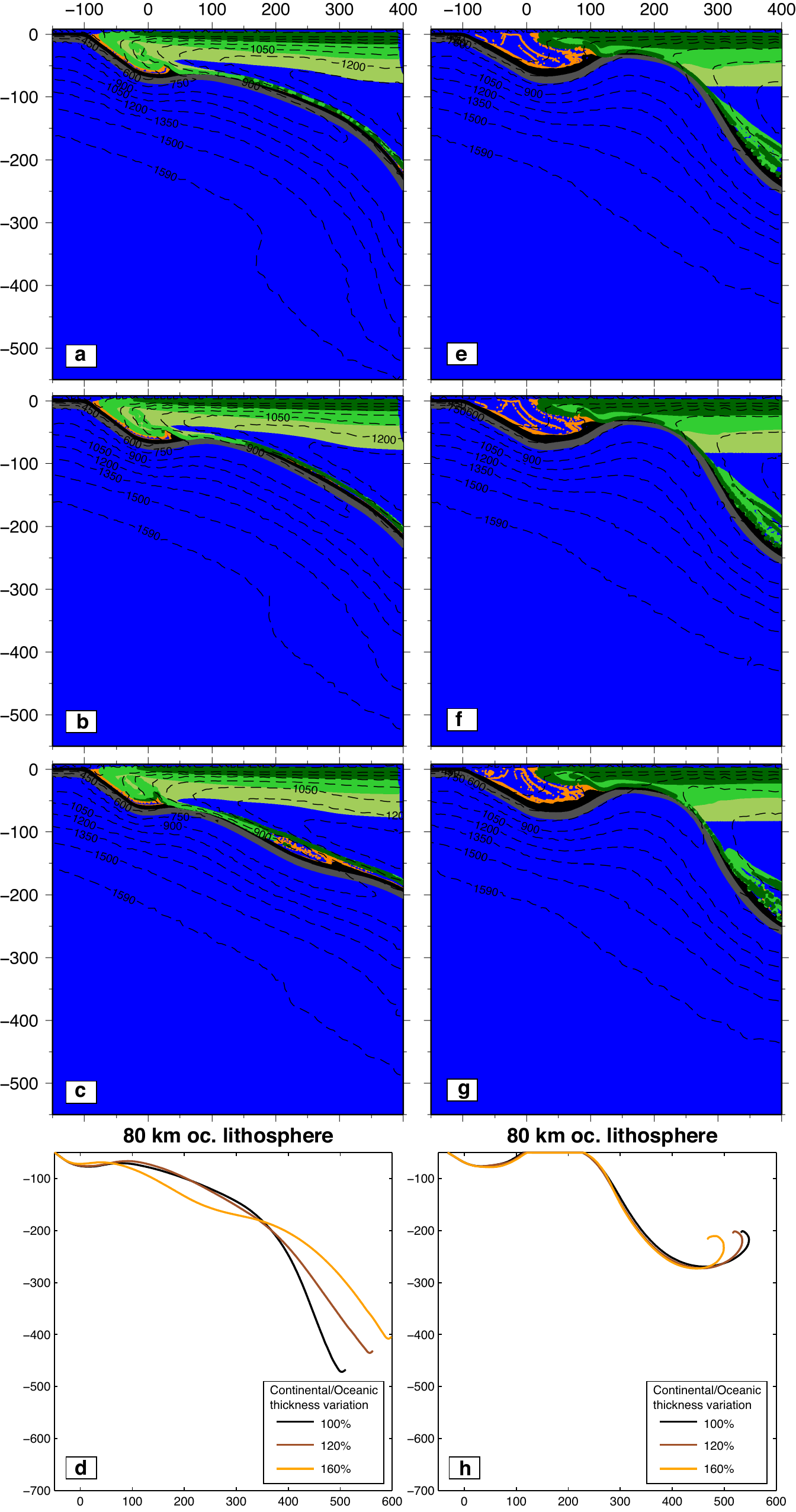}
\caption{Simulation results obtained for 80 km thick oceanic lithosphere and for 3 different thermal states of the overriding plate (100\% (a-e), 120\% (b-f), and 160\% (c-g) of the thickness of the oceanic lithosphere. In panels a, b and c the results for an imposed shallow dip of $37^o$ are shown (see legend of Fig. 2). In panel d the traces of the slab top obtained starting from 50 km depth for each performed simulation, after 30 Ma since the beginning of the subduction, are displayed. In panels e, f and g the results for an imposed shallow dip of $30^o$ are shown (see legend of Fig. 2). In panel h the traces of the slab top obtained starting from 50 km depth for each performed simulation, after 30 Ma since the beginning of the subduction, are displayed.}
\end{figure}

\begin{figure}
\noindent\includegraphics[width=20pc]{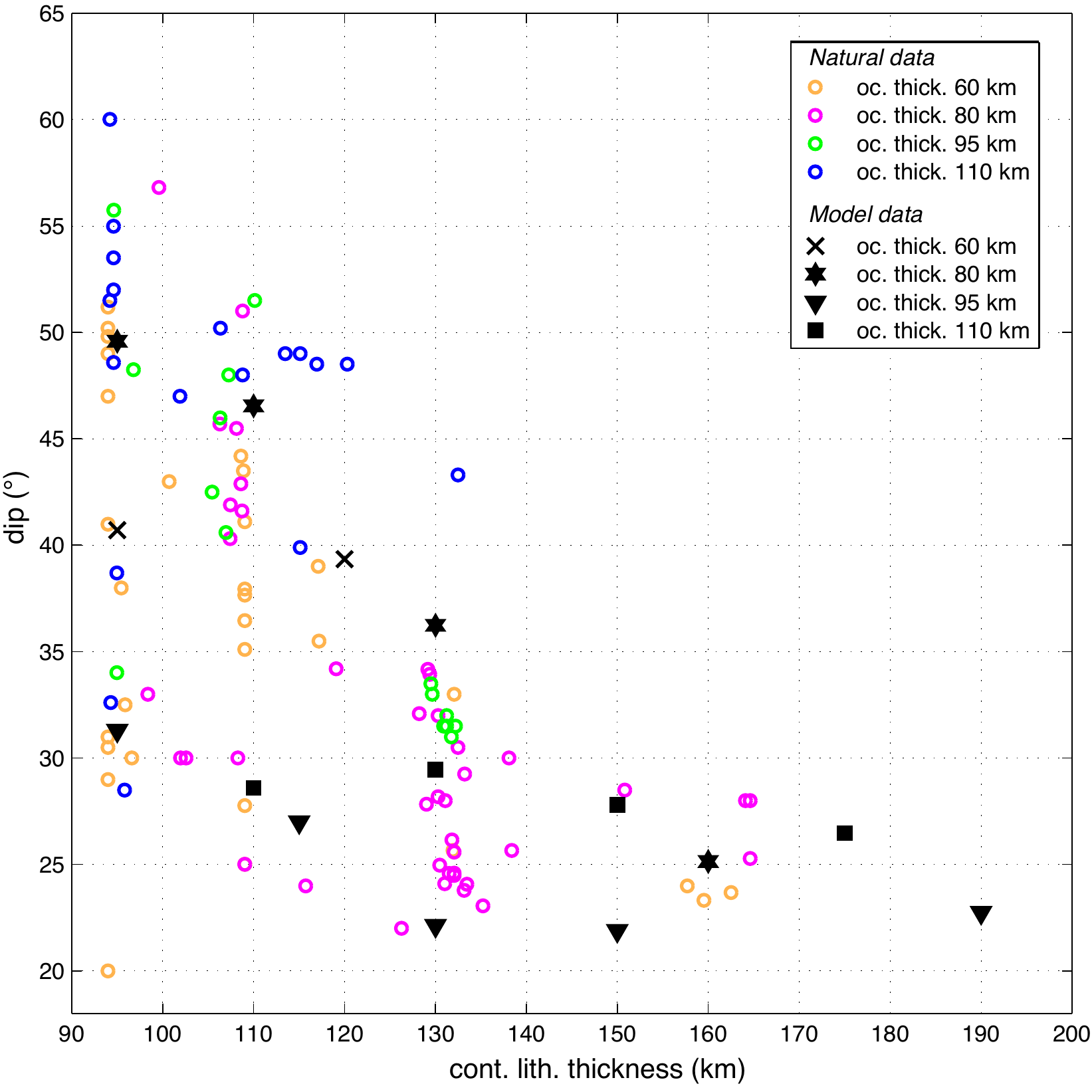}
\caption{Slab dip vs thickness of the continental lithosphere: comparison between natural data (empty circles) and model predictions (black symbols) grouped for 4 different thicknesses of the oceanic lithosphere (see the legend).}
\end{figure}

\end{document}